\documentclass[useAMS,usenatbib]{mn2e}

\usepackage{graphicx,amssymb,color,amsmath}

\def\aap{A\&A}                
\def\apj{ApJ}                 
\def\apjl{ApJ}                
\def\araa{ARA\&A}             
\def\mnras{MNRAS}             
\def\nat{Nature}              

\definecolor{DarkOrchid4}{rgb}{0.41,0.13,0.55}

\title[Crystalline silicates in starburst galaxies]{The crystalline fraction of interstellar silicates in starburst galaxies}

\author[F. Kemper, A. J. Markwick and Paul M. Woods]{F. Kemper\thanks{E-mail:
ciskakemper@fastmail.net},$^{1,2}$ A. J.
Markwick$^1$ and Paul M. Woods$^1$\\
$^1$Jodrell Bank Centre for Astrophysics, Alan Turing
Building, School of Physics and Astronomy, The University of
Manchester,\\ Oxford Road, Manchester, M13 9PL, UK\\
$^2$Academia
  Sinica Institute of Astronomy and Astrophysics, P.O. Box 23-141,
  Taipei 10617, Taiwan, R.O.C.\\
}
\begin{document}



\maketitle


\begin{abstract}
  We present a model using the evolution of the stellar population in
  a starburst galaxy to predict the crystallinity of the silicates in
  the interstellar medium of this galaxy. We take into account dust
  production in stellar ejecta, and amorphisation and destruction in
  the interstellar medium and find that a detectable amount of
  crystalline silicates may be formed, {particularly at high
    star formation rates, and in case supernovae are efficient dust
    producers}.  We discuss the effect of dust destruction and
  amorphisation by supernovae, and the {effect of a} low dust-production
  efficiency by supernovae, and find that when taking this into
  account, crystallinity in the interstellar medium becomes hard to
  detect.  Levels of 6.5--13\% crystallinity in the interstellar
  medium of starburst galaxies have been observed and thus we conclude
  that not all these crystalline silicates can be of stellar origin,
  and an additional source of crystalline silicates {associated
    with the Active Galactic Nucleus} must be present.

\end{abstract}


\section{Introduction}

Silicates are among the most commonly-found dust species in the
interstellar medium {(ISM)} of galaxies. Their presence is established
through the detection of the mid-infrared resonances due to the Si-O
stretching and the O-Si-O bending mode at 9.7 and 18 $\mu$m
respectively.  In galaxies, these bands are seen in absorption
\citep{GFM_75_HII,SLT_00_galaxies}, as well as in emission
\citep{2005A&A...436L...5S,2005ApJ...625L..75H}. Most of these
silicates show the broad resonances characteristic of amorphous
silicates, i.e.,~silicates showing a large degree of lattice defects,
and it is generally assumed that silicates in the ISM are
predominantly amorphous.  In particular, the degree of crystallinity
$x$, defined as the mass fraction of silicates that is crystalline,
$x=M_X/(M_X+M_A)$, in the Galactic diffuse ISM is found to be of the
order of 1\% \citep{KVT_04_GC,KVT_05_erratum}. This contrasts sharply
with the much higher degree of crystallinity seen in silicates in the
circumstellar environments of pre- and post-main-sequence stars
\citep[see e.g.][and references herein]{MK_05_2005}. Generally
speaking, the galactic cycle of dust starts with its formation in
evolved stars, followed by processing in the ISM and eventually ends
with incorporation in stars and planets during star formation. The
silicates observed around Asymptotic Giant Branch (AGB) stars can have
significant crystalline fractions, in particular for the high
mass-loss rate OH/IR stars \citep[up to
$\sim$20\%][]{SKB_99_ohir,KWD_01_xsilvsmdot,2010A&A...516A..86D}. For
lower mass-loss rate AGB stars, such as Miras, the crystallinity is
not well established, but is consistent with a value that does not
vary with mass-loss rate \citep{KWD_01_xsilvsmdot}.  For more massive
stars, such as Red Supergiants (RSGs), the crystallinity of the
silicates in the stellar ejecta is not well known, although isolated
studies report high crystalline fractions
\citep[e.g.~][]{MWT_99_afgl4106}, {while low crystalline
  fractions seem to be more common \citep{2009A&A...498..127V}.
  \citet{2006ApJ...638..759S} adopt a crystallinity of 15\% for
  RSGs and Luminous Blue Variables.}

In contrast to the low crystallinity in the Galactic ISM, significant
amounts of crystalline silicates have been detected in the infrared
spectra of Ultraluminous Infrared Galaxies
\citep[ULIRGs;][]{2006ApJ...638..759S}. In a sample of 77 ULIRGs, 12
were found to show crystallinity, with crystalline-to-amorphous
silicate mass ratios ranging from 0.07 to 0.15, corresponding to
crystallinities of 6.5\% to 13\%. 

The degree of crystallinity of a population of silicate grains
provides a record of the processing history of those grains \citep[see
e.g.~][]{MK_05_2005}. A high crystalline fraction points to a
relatively high formation or processing temperature ($\sim 1000 -
1500$ K), while a large amorphous fraction indicates that the
population of grains has undergone the damaging effects of cosmic ray
hits \citep{2007ApJ...662..372B}, grain-grain collisions or atomic
impacts in shocks
\citep{DCL_01_He+,BSB_03_amorphisation,JFS_03_bombardment}; or that it
is formed at lower temperatures.  Crystalline silicates may thus form
in dense circumstellar environments, where the vicinity of the central
star provides the required heating. The mass-loss processes of evolved
stars subsequently spread these crystalline silicates into the ISM.
The fact that the silicates in the ISM of the Milky Way are almost
entirely amorphous \citep{KVT_04_GC} suggests that the amorphisation
processes in the ISM are more important than the injection of fresh
crystalline silicates into the interstellar reservoir by evolved
stars. For the Galaxy, an amorphisation time scale of 40 Myr has been
derived from observations \citep{KVT_04_GC,KVT_05_erratum}, which is
close to the experimental value of 70 Myr \citep{2007ApJ...662..372B}.

{\citet{2006ApJ...638..759S}
  argue that in starburst galaxies ultraviolet (UV) photons from
  either the Active Galactic Nuclei (AGN) or massive stars are the
  only potential sources of UV photons to anneal or form the
  crystalline silicate mass observed. Crystallisation due to UV
  photons originating from the AGN is dismissed by
  \citet{2006ApJ...638..759S}, because of the observed lack of crystalline
  silicates in the inner 2 pc of NGC 1068 \citep{2004Natur.429...47J_short},
  and the fact that the crystalline silicates are only seen in
  absorption, and can therefore not be very warm.
  \citet{2006ApJ...638..759S}} hypothesise that the crystalline
silicates {must be} produced by massive stars originating from
the starburst.
In this paper, we will investigate the {viability} of the
build-up of crystalline silicates {due to the} starburst
{activity in ULIRGS as proposed by
  \citet{2006ApJ...638..759S}}, and compare the crystalline fraction
to the levels observed by {these authors}. {The
  alternative formation of crystalline silicates due to AGN activity
  will be the subject of a separate future study. This future work
  will be based on the predicted formation of dust in the quasar wind
  rising from the accretion disk \citep{2002ApJ...567L.107E}, where
  conditions may be similar to those present in AGB star winds, a class of
  very efficient dust producers.  Crystalline silicates may form in
  quasar winds if the conditions are right.  Indeed, we have already
  observed crystalline silicates in the quasar wind of PG 2112+059
  \citep{2007ApJ...668L.107M}.}

\section{The model}

\subsection{Injection of mass in the ISM of starburst galaxies}
\label{sec:injection}

When a starburst occurs, a large fraction of the gas available in
molecular clouds will be triggered into forming stars.

Essentially following the method described by
\citet{1980FCPh....5..287T}, we assume that the burst of star
formation occurs at a constant rate $\psi(t) = \psi_0$ for $t_0 < t
<t_1$, and that the star formation rate before $t_0$ and after $t_1$
is negligible.  The distribution of stars over the mass spectrum
follows the initial mass function (IMF) $\phi(m) = N_0 m^{-(1+a)}$. We
adopt the IMF as described by \citet{2001MNRAS.322..231K}, who derive
a value of $a=1.3$ for $M>1\,M_\odot$, which is not too dissimilar
from the original Salpeter IMF \citep[$a=1.35$ for the entire mass
range;][]{S_55_IMF}. About 51\% of the mass in a star formation event
is contained in the stars more massive than the Sun
\citep{2001MNRAS.322..231K}, from which we can calculate $N_0$ by
solving

\begin{equation}
\int_{m=1}^{\infty} \phi(m) dm = N_0 \int_{1}^{\infty}
m^{-(1+a)} dm = 0.51 \, .
\end{equation}

We find that $N_0 = 0.663$ satisfies this equation. Since the duration
of the starburst will be much shorter than the life time of the Sun
($\tau_\odot$), we do not need to consider mass-loss for low mass
stars, so the exact shape of the ISM below $M < 1\,M_\odot$ is
irrelevant. Naturally, setting the upper integration boundary to
$m=\infty$ results in unrealistically high stellar masses, but since
the IMF is a very steep power-law, the numerical contribution from
hypothetical stars with $M>120\,M_\odot$ is negligible.

The time that a star spends on the Main Sequence is approximated by
$\tau_{\mathrm{MS}} \propto \frac{M}{L}$, while the mass-luminosity
relation states that $L \propto M^\eta$, with $\eta \approx 3.35$ for
stars with $1\,M_{\odot}<M<40\,M_{\odot}$
\citep[e.g.][]{KW_90_stellar}. Thus we find for the {lifetime} of
a star on the Main Sequence -- using solar units, e.g.~$m(M_\odot)$
and $t(\tau_\odot)$ -- that $ \tau_{m} = m^{1-\eta}$.

It is possible to set up an equation for the ejection rate $E(t)$ at
which gas is being expelled from a population of stars formed during a
starburst in a galaxy \citep{1980FCPh....5..287T}.  Assuming the
mass{-}loss occurs {instantaneously} after a lifetime
$\tau_m$ on the {Main Sequence}, a star formed at time
$(t-\tau_m)$ ejects its shell at time $t$. In order to obtain the
ejection rate, one needs to integrate over all masses $m$ larger than
$m_t$, the stellar mass corresponding to {lifetime} $\tau_m =
t$. Thus, the ejection rate is

\begin{equation}
E(t) = \int_{m_t}^{\infty} (m-w_m) \psi(t-\tau_m) \phi(m) dm
\label{eq:ej}
\end{equation}

\noindent where $w_m$ {is} the mass of the remnant after the 
mass loss. By substituting $\tau_m = m^{1-\eta}$ and
$m_t=t^{\frac{1}{1-\eta}}$ and using the initial mass function and our assumption
that the star formation rate is constant at a value of $\psi_0$ during the
burst, the
ejection rate $E(t)$ can be written as

\begin{equation}
E(t) = \int_{m=t^{\frac{1}{1-\eta}}}^{\infty} (m-w_m)
\psi_0 N_0 m^{-(1+a)} dm \, .
\label{eq:int}
\end{equation}

\subsection{The remnant mass function}

We adopt the strategy utilised by \citet{2009MNRAS.394.1529D} to
describe the remnant mass function, $w_m$.  It slices the stars with
$m \geq 1$ into three categories: those that leave white dwarfs ($m <
8$), those that leave neutron stars ($8 \leq m <25$) and those that
leave black holes ($m \geq 25$) as their stellar remnants.
\citet{2009MNRAS.394.1529D} arrive at the following expressions for
the remnant mass of the form $w_m = f_w m + w_{m_0}$, with $f_w$ and
$w_{m_0}$ determined in each category to be:

\begin{equation}
w_m = \begin{cases}
0.109\, m + 0.394 & \mathrm{for} \qquad 1 \leq m < 8   \\
1.35                          & \mathrm{for} \qquad 8 \leq m < 25    \\
0.1\, m           & \mathrm{for} \qquad 25 \leq m
\end{cases}
\label{eq:remnantmass}
\end{equation} 

\subsection{A starburst of finite duration}

 In order to calculate the ejection rate, we can distinguish
two different eras: during the starburst ($0<t<\sigma$) and after the
starburst ($t>\sigma$). Within each era, we can subdivide into three 
different regimes, corresponding to the lifetimes of the stellar mass
ranges described by \citet{2009MNRAS.394.1529D}, defined by the boundaries
 $\tau_8=8^{1-\eta} \tau_\odot = 7.5 \cdot 10^{-3} \,
\tau_\odot$ and $\tau_{25}=25^{1-\eta} \tau_\odot = 5.2 \cdot 10^{-4} \,
\tau_\odot$. We assume that $\sigma > \tau_8$. 

\subsubsection{During the starburst: $0<t<\sigma$}

\begin{itemize}

\item Phase I: $ t < \tau_{25}$:

Only the most massive stars ($m>25$) have started losing mass, and for these
objects the remnant mass is given by $w_m = 0.1\, m$ (Eq.~\ref{eq:remnantmass}), and thus Eq.~\ref{eq:int} becomes

\begin{equation}
E(t) = \int_{m=m_t}^{\infty} m(1-f_w) \psi_0 N_0 m^{-(1+a)} dm \label{eq:ejduring1}
\end{equation}

\noindent which simplifies to

\begin{equation}
E(t) = - \psi_0 N_0 (1-f_w) \Bigg(\frac{1}{1-a} {m_t}^{1-a} \Bigg)
\end{equation}

\noindent with $N_0 =0.663$ and $f_w = 0.1$, and $m_t = (t/\tau_\odot)^{{1}/{1-\eta}}$.

\item Phase II: $\tau_{25} < t < \tau_8$:

  During this phase, two components contribute to the ejection rate: $E(t) = E_1 + E_2(t)$, with $E_1$ a constant value due
  to the ejection rate by the most massive stars ($m > 25$), and
  $E_2(t)$ an increasing component due to the ejection by intermediate
  mass stars ($8<m<25$). To obtain $E_1$, we integrate
  Eq.~\ref{eq:ejduring1} between $m=25$ and $\infty$, and find that

\begin{equation}
E_1 = -(1-f_w) \psi_0 N_0 \frac{1}{1-a} 25^{1-a} \, 
\label{eq:E1}
\end{equation}

\noindent which, for $f_w = 0.1$, $N_0 = 0.663$, and $a=1.3$ results in $E_1 = 0.76\, \psi_0$.

To obtain $E_2(t)$, we integrate

\begin{equation}
E_2(t) = \int_{m=m_t}^{25} (m-w_m) \psi_0 N_0 m^{-(1+a)} dm \label{eq:ejduring2}
\end{equation}

\noindent and find

\begin{equation}
E_2(t) = \psi_0 N_0 \Bigg( 25^{-a} \Big( \frac{25}{1-a} + \frac{w_m}{a} \Big) - m_t^{-a} \Big( \frac{m_t}{1-a} + \frac{w_m}{a} \Big) \Bigg) 
\end{equation}

\noindent with $w_m = 1.35$.

\item Phase III: $\tau_8 < t < \sigma$, with $\sigma < \tau_\odot$:

During this phase, the ejection rate can be written as $E(t) = E_1 + E_2 +E_3(t)$ with, $E_1$ given by Eq.~\ref{eq:E1}, and $E_2$ obtained by substituting $m_t=8$ into Eq.~\ref{eq:ejduring2}. This yields

\begin{equation}
E_2 = \psi_0 N_0 \Bigg( 25^{-a} \Big( \frac{25}{1-a} + \frac{w_m}{a} \Big) - 8^{-a} \Big( \frac{8}{1-a} + \frac{w_m}{a} \Big) \Bigg) 
\label{eq:E2}
\end{equation}

\noindent which, for $w_m = 1.35$ gives $E_2 = 0.47\, \psi_0$. 

The ejection rate by low mass stars ($m<8$) $E_3(t)$ can be obtained from Eq.~\ref{eq:int}:

\begin{equation}
E_3(t) = \int_{m=m_t=(t/\tau_\odot)^{\frac{1}{1-\eta}}}^{8} (m-w_m) \psi_0 N_0 m^{-(1+a)} dm\\
\end{equation}
\noindent with $w_m = f_w m + w_{m_0}$. This can be rewritten as
\begin{align}
E_3(t) = \psi_0 N_0 \Bigg[ 8^{1-a} \Bigg( \frac{1-f_w}{1-a} + \frac{w_{m_0}}{a} \cdot \frac{1}{8} \Bigg) \nonumber\\- {m_t}^{1-a} \Bigg( \frac{1-f_w}{1-a} + \frac{w_{m_0}}{a} \cdot \frac{1}{m_t}  \Bigg) \Bigg] \, .
\label{eq:ejduring3}
\end{align}

\end{itemize}

\subsubsection{After the starburst: $t>\sigma$}

When a starburst of duration $\sigma$ has stopped, there are no longer
any stars with $m>((t-\sigma)/\tau_\odot)^{\frac{1}{1-\eta}}$ in the
galaxy, thus constraining the upper limit of the integration to obtain the
ejection rate.

\begin{itemize}
\item Phase IV: $t - \sigma < \tau_{25}$:

  Shortly after the starburst has stopped stars with $m > 25$ still
  exist, and we can write for the ejection rate: $E(t) = E'_1(t) + E_2
  + E_3(t)$. $E_2$ and $E_3(t)$ are given by Eqs.~\ref{eq:E2} and
  \ref{eq:ejduring3}.  For $E'_1(t)$ we can write, with
  $m_{t-\sigma}=((t-\sigma)/\tau_\odot)^{1/1-\eta}$,

\begin{equation}
E'_1(t) = \int_{m=25}^{m_{t-\sigma}} m(1-f_w) \psi_0 N_0 m^{-(1+a)} dm
\end{equation}

\noindent which we can rewrite as:

\begin{equation}
E(t) = \psi_0 N_0 (1-f_w) \Bigg[\frac{1}{1-a} \Bigg\{{m_{t-\sigma}}^{1-a} - 25^{1-a} \Bigg\} \Bigg] \, .
\label{eq:ejafter4}
\end{equation}

\item Phase V: $\tau_{25} < t - \sigma < \tau_8$.

  In this phase, high-mass stars ($m>25$) are extinct and no longer
  contribute to the ejection rate, which has now become $E(t) =
  E'_2(t) + E_3(t)$, with $E_3(t)$ again equal to the result of
  Eq.~\ref{eq:ejduring3}. For $E'_2(t)$ we can write

\begin{align}
E'_2(t) = \int_{m=8}^{m_{t-\sigma}} (m-w_m) \psi_0 N_0 m^{-(1+a)} dm \nonumber\\
= \psi_0 N_0 \Bigg[ {m_{t-\sigma}}^{-a} \Bigg( \frac{m_{t-\sigma}}{1-a} +\frac{w_m}{a} \Bigg) \nonumber\\ - 8^{-a} \Bigg( \frac{8}{1-a} + \frac{w_m}{a} \Bigg) \Bigg] \, 
\label{eq:ejafter5}
\end{align}

\noindent with  $m_{t-\sigma}=((t-\sigma)/\tau_\odot)^{1/1-\eta}$. 

\item Phase VI: $t - \sigma > \tau_8$

At this late time after the starburst only
low-mass stars with $m<8$ are still in existence, and thus 

\begin{align}
E(t) = E'_3(t) \nonumber\\
= \psi_0 N_0 \Bigg[ {m_{t-\sigma}}^{1-a} \Bigg( \frac{1-f_w}{1-a} + \frac{w_{m_0}}{a} \cdot \frac{1}{m_{t-\sigma}} \Bigg) \nonumber\\- {m_t}^{1-a} \Bigg( \frac{1-f_w}{1-a} + \frac{w_{m_0}}{a} \cdot \frac{1}{m_t}  \Bigg) \Bigg] \, .
\label{eq:ejafter6}
\end{align}
 
\end{itemize}

\subsection{Mass ejection rates and dust production}
\label{sec:dustprod}

\begin{figure}
\includegraphics[width=8.4cm]{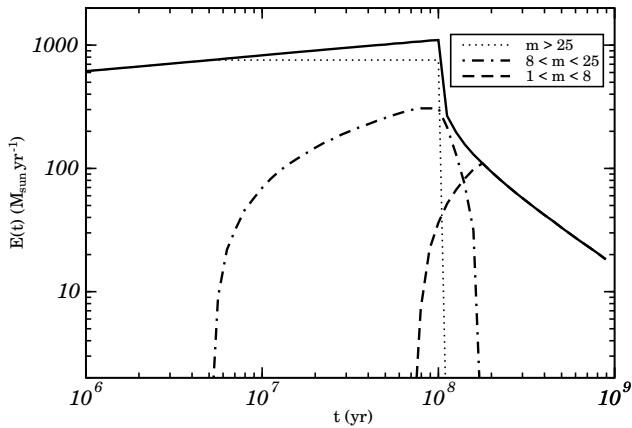}
\caption{ Stellar mass ejection rate into the ISM of a starburst
  galaxy with a star formation rate $\psi = 1000\, M_\odot$ yr$^{-1}$,
  as a function of time after the beginning of the starburst (solid
  line). The duration of the starburst is 100 Myr. The dotted,
  dash-dotted and dashed lines indicate the contributions by high,
  intermediate and low mass stars respectively.}
\label{fig:ejrate}
\end{figure}

In Fig.~\ref{fig:ejrate} the resulting ejection rate $E(t)$ throughout
phases I--VI is shown for a starburst of 1000 $M_\odot$ yr$^{-1}$ and
a duration of 100 Myr, as well as the ejection by high, intermediate
and low mass stars specifically. It is clear that during the burst of
star formation supernovae ejecta from the most massive stars ($m>25$)
dominate the mass return to the ISM, but that after the starburst has
ended, intermediate- and low-mass stars are important.

\begin{figure}
\includegraphics[width=8.4cm]{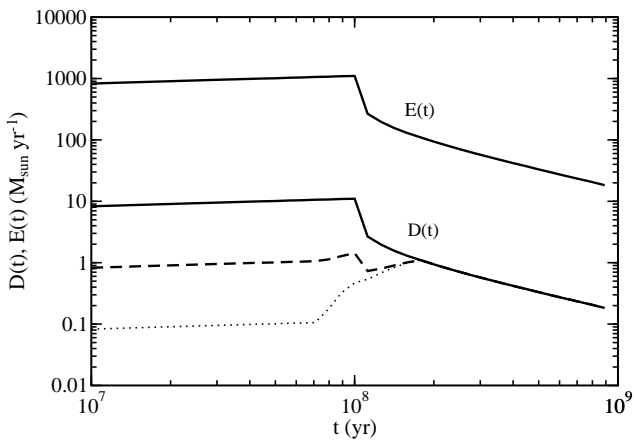}
\caption{ Dust production rate $D(t)$ in a starburst galaxy, compared
  to the stellar mass ejection rate $E(t)$, both indicated by solid
  lines. The dust-to-gas ratio $d$ is assumed to be 0.01 in all
  ejecta. The dashed and dotted lines show the dust production rates
  in cases where supernovae produce dust at 10 or 100 times lower
  efficiencies, respectively.}
\label{fig:dustejrate}
\end{figure}

For each of the regimes, the dust production rate $D(t)$ is related to
the mass ejection rate by $D(t) = E(t) d$, in which $d$ is the
dust-to-gas ratio. In the simplest scenario we assume that low,
intermediate and high mass stars are all equally efficient dust
producers, for which we use the commonly-used value of $d=0.01$. This
gives rise to the solid line labelled $D(t)$ in
Fig.~\ref{fig:dustejrate}.  The dust-production efficiency of
supernovae is a hot topic in research because of its implications for
the presence of dust at high redshift, and the consensus seems to be
that dust production in supernova ejecta is not as efficient as in the
outflows of lower mass stars
\citep[e.g.~][]{2006Sci...313..196S,2010arXiv1005.4682F}. We have thus
also considered values of 0.001 and 0.0001 for $d$, for the ejecta of
high and intermediate mass stars ($m>8$), all of which produce
supernovae.  The resulting $D(t)$ is shown in
Fig.~\ref{fig:dustejrate}, by the dashed and dotted lines
respectively.

\subsection{The dust budget in the ISM of a galaxy}

In our model, we will consider two separate dust reservoirs to be
present in a galaxy: a crystalline silicate reservoir with a mass
$M_{\mathrm{X}}$ and an amorphous silicate reservoir with a mass
$M_{\mathrm{A}}$.  The injection of new silicate
material $D(t)$, as described in Sect.~\ref{sec:dustprod}, affects
both reservoirs, although not in equal amounts. We use $x_\star$ to
denote the fraction of silicate mass in crystalline form produced by
stars. Another possible source of silicates is dust formation in the
ISM itself, which can be generally expressed as $F_\mathrm{A}(t)$ and
$F_\mathrm{X}(t)$ for amorphous and crystalline silicates
respectively.

We include the transitions between both reservoirs due to annealing or
destruction of the lattice structure, which occur at a rate
{$k_1$ for the amorphisation of crystalline material and $k_2$
  for the crystallisation of amorphous silicates.}

Finally, the processes that remove material from both dust reservoirs
include grain destruction by supernova shocks and cosmic ray hits, but
also astration and incorporation in planet{-}forming disks. We
summarise these terms as dust destruction, occurring at rates of $k_3$
and $k_4$ for crystalline and amorphous silicates respectively.

Following Eq.~5 from \citet{KVT_04_GC}, these rates can be
written as a set of coupled differential equations:

\begin{equation}
\label{eq:diff}
\left\{ 
\begin{array}{l} 
\frac{dM_{\mathrm{X}}}{dt} = x_{\ast}D(t) - k_1 M_{\mathrm{X}} + k_2
M_{\mathrm{A}} - k_3 M_{\mathrm{X}} + F_\mathrm{X}(t)\\
\frac{dM_{\mathrm{A}}}{dt} = (1 - x_{\ast})D(t) + k_1 M_{\mathrm{X}} -
k_2 M_{\mathrm{A}} - k_4 M_{\mathrm{A}} +F_\mathrm{A}(t) \, . \\
\end{array}  \right.
\end{equation}

\noindent In interstellar conditions, the crystallisation rate $k_2$ will be
negligible, as annealing requires the grains to be heated to $\gtrsim
1\,000$ K, which is unlikely to occur. We also assume that the
destruction rates for both types of silicates are equal: {$k_3 =
  k_4 = 2 \times 10^{-9}$\,yr$^{-1}$ \citep{KVT_04_GC}}, and that dust
formation in the ISM can be ignored: $F_{\mathrm{X}}(t) =
F_{\mathrm{A}}(t) = 0$. {For the amorphisation rate of
  crystalline silicates we take the rate $k_1 = 2.5 \times
  10^{-8}$\,yr$^{-1}$, in accordance with \citet{KVT_05_erratum}}.

Eq.~\ref{eq:diff} allows us to investigate the time dependence of the
crystalline and amorphous silicate masses, and compare the results
with the observational constraints \citep{2006ApJ...638..759S}.
Instantaneous mixing of the silicates is assumed.

\subsection{Fitting parameters}

The values for the fixed parameters and the fitting ranges for the
free parameters required to reproduce the observed crystallinities are
set as follows.  The star formation rate $\psi_0$ is allowed to vary
between 10--1000 $M_\odot$ yr$^{-1}$. The total initial dust mass in a
galaxy is varied between 10$^7$--10$^8$ $M_\odot$, with an initial
crystalline fraction $x_{\mathrm{ISM}} \leq 0.01$. It should be noted
that both the initial dust mass and the star formation rate are
related to the gas mass in the galaxy, and both will go up for a
larger gas reservoir. The crystallinity in the stellar ejecta $x_\ast$
is taken to be 0.10--0.20.  The duration of the starburst is allowed
to run until 0.01 $\tau_\odot$, or 100 Myr, which is well beyond the
commonly accepted value of 5--10 Myr for the duration of starburst,
although recent work on dwarf galaxies suggest that starburst may last
a few 100 Myr \citep{2010arXiv1009.2940M}.
  
\section{Results}
\label{sec:results}

\begin{figure}
\includegraphics[width=8cm]{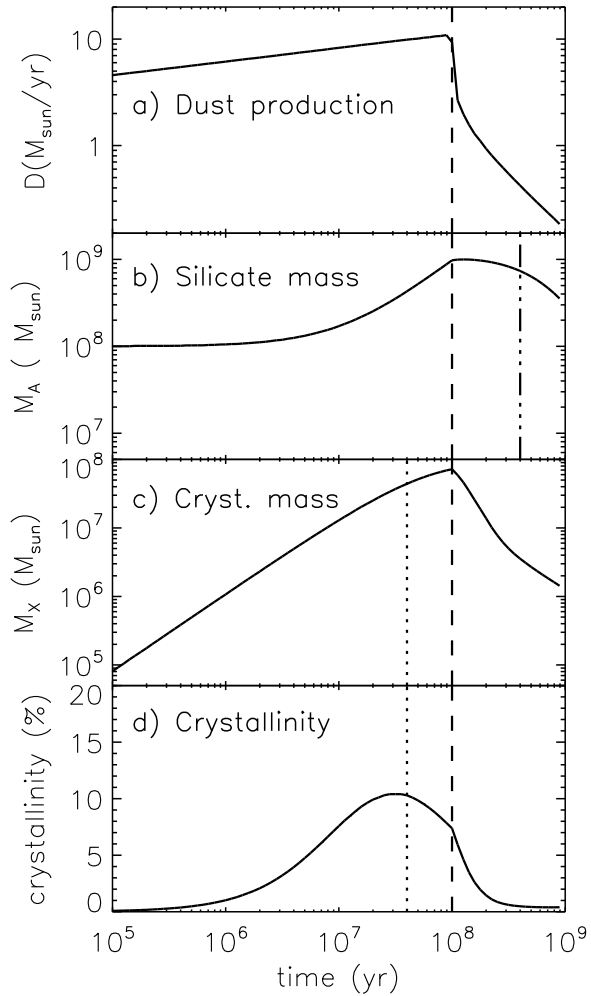}
\caption{Results from model calculations, with the initial
  interstellar silicate mass $M_X + M_A = 10^{8} \, M_\odot$, the star
  formation rate $\psi_0 = 1000 \, M_\odot$ yr$^{-1}$, the stellar
  crystallinity $x_\ast = 0.2$ and the dust-to-gas ratio $d=0.01$ for
  all stars. Panel {\bf a)} displays the total injection rate of
  stellar dust into the ISM as a function of time after the beginning
  of the starburst.  In panel {\bf b)} the total silicate mass in the
  galaxy is shown. Panel {\bf c)} shows the crystalline silicate mass
  in solar masses, and finally, in panel {\bf d)} we show the
  crystallinity, defined as the crystalline silicate mass divided by
  the total silicate mass. The dashed line indicates the end of the
  starburst $t_1 = 10^8$ yrs. The dashed-triple-dotted and the dotted
  lines indicate destruction and amorphisation time scales in the ISM
  respectively. }
\label{fig:xev}
\end{figure}

We have numerically evolved the system using Eq.~\ref{eq:diff}. The
results of a run with an initial dust mass
$M_{\mathrm{X}}+M_{\mathrm{A}} = 10^8\,M_\odot$, star formation rate
$\psi_0 = 1000\,M_\odot$ yr$^{-1}$ and crystallinity $x_\ast = 0.2$
for the silicates in the stellar ejecta in Fig.~\ref{fig:xev}. In this
case supernovae are considered equally efficient dust producers as low
mass stars, with $d = 0.01$ over the entire stellar mass range.  Panel
{\bf a)} shows the silicate ejection rate throughout time for a
starburst of 0.01 $\tau_\odot$ in duration. Due to the contribution of
these ejecta, the total silicate mass $M_{\mathrm{X}}+M_{\mathrm{A}}$
in the galaxy increases (panel {\bf b)}), and levels off after the end
of the starburst (dashed line). The dash-triple-dotted line indicates
the destruction time scale of 400 Myr, which is due to rates $k_3$ and
$k_4$. At this point the total dust mass has decreased significantly
from its peak during the starburst.  The mass $M_{\mathrm{X}}$
contained in crystalline silicates (panel {\bf c)}) closely follows
the ejection rate, since the amorphisation time scale is short (40
Myr, based on $k_1$, indicated with a dotted line). The crystallinity
in the ISM $x_{\mathrm{ISM}} =
M_{\mathrm{X}}/(M_{\mathrm{X}}+M_{\mathrm{A}})$ \citep{KVT_04_GC} is
plotted in panel {\bf d)} and appears to peak before the amorphisation
time scale has elapsed, at around 20 Myr. A significant level of
crystallinity may build up for a brief period of time. In this
particular example, we find levels of $x_{\mathrm{ISM}} \approx
10-11$\%.

The definition of crystallinity used here is based on the work
presented in an earlier paper \citep{KVT_04_GC}, and differs from the
crystalline-to-amorphous ratio $N_{\mathrm{cr}}/N_{\mathrm{am}}$
measured by \citet{2006ApJ...638..759S}. Their ratios of 0.07--0.15
therefore translate to crystallinities $x_{\mathrm{ISM}}$ of 6.5\% --
13\% for the interstellar silicates in these ULIRGs, experiencing
starbursts.  The peak crystallinity in our calculation of $\sim
10-11$\% falls within this range.

\begin{figure}
\includegraphics[width=8cm]{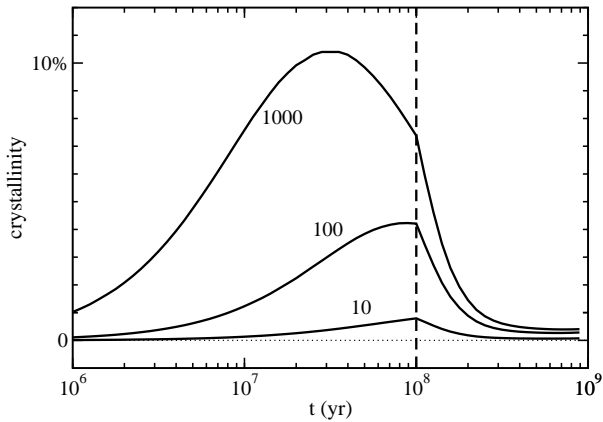}
\caption{Crystallinity in models with star formation rates $\psi_0$ of
  1000, 100 and 10 $M_\odot$ yr$^{-1}$ (solid lines). The other model
  parameters are set as described in the beginning of
  Sect.~\ref{sec:results}. The dashed line indicates the end of the
  starburst.  }
\label{fig:sfr}
\end{figure}

However, a star formation rate of $\psi_0 = 1000 \, M_\odot$ yr$^{-1}$
is on the high end of the range seen in starburst galaxies
\citep[e.g.~][]{2008ApJ...686..127W}, and values of 10--100 seem to be
more common \citep[e.g.~][]{1998ARA&A..36..189K}. When these lower
values are used in the model calculations, the crystallinity of the
silicates in the ISM builds up more slowly (Fig.~\ref{fig:sfr}), and peaks
at values below the measurements by \citet{2006ApJ...638..759S}.

\subsection{The effect of supernovae}

\begin{figure}
\includegraphics[width=8cm]{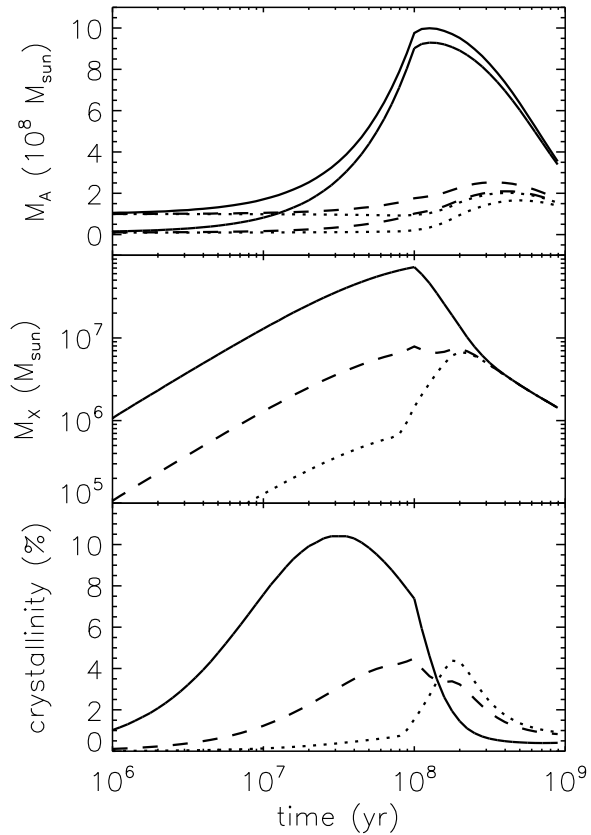}
\caption{{The amorphous and crystalline dust masses in a
    starburst over time, for three different SN dust production
    efficiencies: $d=0.01$ (solid lines), $d=0.001$ (dashed lines) and
    $d=0.0001$ (dotted lines). The first panel shows the amorphous
    silicate mass $M_A$ in the ISM of a starburst galaxy, with the two
    curves in each line style representing the initial ISM dust masses
    of $10^7$ $M_\odot$ (lower lines) and $10^8$ $M_\odot$ (upper
    lines). The second panel shows the crystalline silicate mass
    $M_X$, and the third panel the crystalline fraction
    $x=M_X/(M_X+M_A)$ of silicates in the ISM of a starburst galaxy as
    a function of time.  For all models, $\psi_0$ is set to 1000
    $M_\odot$ yr$^{-1}$.}  }
\label{fig:asil}
\end{figure}

As Figs.~\ref{fig:xev} and \ref{fig:sfr} demonstrate, the levels of
crystallinity reached in starburst galaxies with high (1000 $M_\odot$
yr$^{-1}$; Fig.~\ref{fig:xev}) and intermediate ($> 100 M_\odot$
yr$^{-1}$; Fig.~\ref{fig:sfr}) star formation rates are in line with
the range of $x=6.5-13$\% that is observed by
\citet{2006ApJ...638..759S}. However, these calculations were
performed under the assumption that the dust production by supernovae
is efficient, with a dust-to-gas mass ratio $d$ in the stellar ejecta
of 0.01, independent of initial stellar mass. However, when we,
following work by e.g.~
\citet{2006Sci...313..196S} and \citet{2010arXiv1005.4682F} confirming
that supernova are not efficient dust producers, consider a 10 or
100 {times} lower dust formation efficiency in the supernovae ejecta
(affecting all stars with $m>8$) a different picture arises.

{The top two panels of Fig.~\ref{fig:asil}} show the amorphous
and crystalline silicate mass in the starburst galaxy, using values of
$d=0.01$ (solid lines), 0.001 (dashed lines) and 0.0001 (dotted lines)
describing the supernova dust productivity, while the dust production
efficiency of low mass stars ($m<8$) is kept at $d=0.01$. The two sets
of lines in {the top panel of} Fig.~\ref{fig:asil} arise from
different initial dust masses in the ISM (see caption).  These plots
show that the overall dust production dramatically decreases, and that
the peak in dust mass shifts to a later time.  The crystalline
silicate mass is even 1 or 2 orders of magnitude lower during the
starburst phase, and is no longer dependent on {the supernova dust
productivity} $~20$ Myr
after the starburst. This is due to the fact that the amorphisation
time scale is very short, and all crystalline silicates present at
this time are freshly produced by low mass stars.

The crystallinity $x$, in {the bottom panel of}
Fig.~\ref{fig:asil} shown for $\psi_0 =1000 \, M_\odot$ yr$^{-1}$ and
an initial dust mass of $10^8$ $M_\odot$, evolves differently with
time for different dust production efficiencies in supernova, however
the width of the curve is determined by the amorphisation rate $k_1$.
At the lowest efficiency we considered, $d=0.0001$ the crystalline
fraction is in fact dominated by the produce of low mass stars, which
peaks after the end of the starburst, in this case. The peak
crystalline fraction is also considerably lower for low dust
production efficiencies, and for this very high star formation rate
only a level of 4\% is reached, just below the lowest values of 6.5\%
observed by \citet{2006ApJ...638..759S}. After 5--10 Myr, the
generally accepted value for the duration of a starburst, only small
amounts of crystallinity have built up, even for the most favourable
set of parameters.

\section{Discussion}

Through constructing an ISM dust evolution model we have been able to
explain the crystallinities observed by \citet{2006ApJ...638..759S} in
a sample of 12 ULIRGs, out of a larger set of 77 analysed. The
observed crystallinities range from 6.5 to 13\%. {We remark,
  however, that the derived crystallinity is high, due to the use of
  the 16.5 $\mu$m forsterite feature.  This is intrinsically one of
  the weaker forsterite resonances \citep{JMD_98_crystalline}, and
  thus a large amount of forsterite is required to explain the optical
  depth in this feature. \citet{2006ApJ...638..759S} chose this
  feature for their analysis because it was the only feature clearly
  observed in all 12 objects. Using Fig.~5 and Eq.~(1) from their
  paper, we estimate that the crystallinity could be a factor of
  $\sim$3 or $\sim$4.5 lower, when using the 18 or 23 $\mu$m features
  respectively.  This should, however, be compared to the lack of a
  crystalline silicate detection in the ISM of the Milky Way
  \citep{KVT_04_GC}.  Additionally, crystallinities in the ULIRG
  sample can only be derived for all sources using the 16.5 $\mu$m
  feature, and we therefore use the results presented by
  \citet{2006ApJ...638..759S} at face value.  }

Using the relatively high values of 1000 $M_\odot$ yr$^{-1}$
for the star formation rate and a crystallinity of 20\% in the stellar
ejecta we achieve interstellar crystallinities of $>10$\%, under the
assumption that supernovae are efficient dust producers, with a dust-to-gas
ratio $d=0.01$ in their ejecta.
As the time-dependent plots show, detectable levels of crystallinity
are only a temporary condition, which is consistent with only seeing
it in part of the {ULIRG} sample studied, with the
timing and the duration of detectable levels depending
on the model parameters. In the case that supernovae are important
dust producers, the crystallinity of the silicates in the ISM peaks
shortly after the starburst on a time scale shorter than the
amorphisation time scale of 40 Myr.  For the parameters used {in our
  fiducial model}, we find that the crystallinity is highest some
10--20 Myr after the starburst {and that crystalline silicates can
  make up $\approx$11\% of the silicate material.}

\subsection{Evaluation of the input parameters}

The input parameters used are on the extreme end of the range
observed, particularly for {the star formation rate} $\psi_0$, 
{the crystallinity} $x_\ast$ {of the silicates in
stellar ejecta} and {the (supernova) dust formation efficiency $d$}. {We also discuss the role of supernovae in general}. 

{First,}
for more typical
star formation rates of $\psi_0 = 10-100$ $M_\odot$ yr$^{-1}$ the
interstellar crystallinity is much less {enhanced} during the starburst
({see} Fig.~\ref{fig:sfr}), making it harder to explain the observed values.

{Second,} the interstellar crystallinity scales linearly with
the crystallinity of stellar ejecta used, where $x_\ast = 20$\% is on
the high side. More typical values are around 10\%
\citep{KWD_01_xsilvsmdot}, although \citet{2010A&A...516A..86D} have
shown that the forsterite abundance (one of the species contributing
to the crystallinity) is closer to 12\%. {Unfortunately,}
{the crystallinity} {can only be measured} in stars with
the densest winds \citep{SKB_99_ohir}; crystallinities up to about
20\% in the more numerous low density winds would remain undetected
\citep{KWD_01_xsilvsmdot}. {Thus, 20\% could be regarded as an
  upper limit to the crystallinity in stellar ejecta, for classes of
  objects where no specific measurements exist.}

\subsubsection{The role of supernovae}

The role of supernovae in the mineralogical
evolution of a galaxy remains uncertain, for a number of reasons.
First, little is known about the mineralogy of supernovae ejecta. The
small number of studies available \citep[e.g.~][]{2009ApJ...700..579R}
do not establish the crystallinity of the produced silicates. Second,
the dust production efficiency $d$ of supernovae has been shown to be
low \citep[e.g.~][]{2006Sci...313..196S,2010arXiv1005.4682F} compared
to lower mass stars. {On the other hand, studies of the Cas A
  and Kepler supernova remnants suggest significant amounts of dust
  may have formed \citep{MDE_03_kepler,DEI_03_SNdust}, although in one
  of these particular studies foreground emission may have been
  interpreted as coming from the supernova remnant itself
  \citep{KBR_04_CasA}. Thus, $d$ may range from 0.0001 to 0.01.}  In
  Fig.~\ref{fig:asil}, the effect of a low dust formation efficiency
  in supernovae on the dust population in a galaxy shows that the
  crystallinity does peak at much later times and at lower values due
  to his effect. Indeed, at a supernova dust production rate {of $d=0.0001$}, the dust production, and thus
  the crystallinity, is dominated by low-mass stars ($m<8$, {with $d=0.01$}), and
  {by} the time the crystallinity peaks these galaxies are no longer
  recognisable as starburst galaxies.

\begin{figure}
\includegraphics[width=8cm]{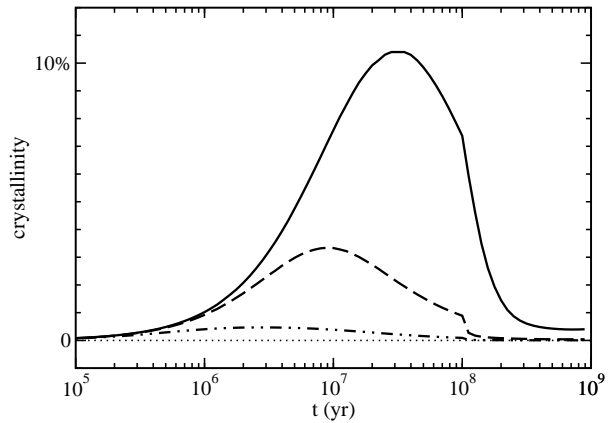}
\caption{Crystalline fraction of silicates $x=M_X/(M_X+M_A)$ in the ISM of a starburst galaxy as a function of time, for three different amorphisation rates $k_1=2.5\cdot 10^{-8}$ (solid line), $k_1=2.5\cdot 10^{-7}$ (dashed line) and $k_1=2.5\cdot 10^{-6}$ yr$^{-1}$ (dash-dotted line). The curves shown are for the initial dust mass of $10^8$ $M_\odot$, and a star formation rate of 1000 $M_\odot$ yr$^{-1}$.
 }
\label{fig:k1}
\end{figure}

Third, the higher supernova rate has its effect on the amorphisation
rate. The values used for $k_1$ are derived for Galactic conditions
but these may be significantly higher in starburst galaxies due to the
enhanced rate of shocks and cosmic ray hits.  We have investigated the
effect of an enhanced supernova rate, by increasing $k_1$ up to a
factor of 100.  Increasing these rates is detrimental to the
crystalline fraction, with an increase of a factor 100 reducing the
crystalline fraction in the ISM to $\ll$1\% (Fig.~\ref{fig:k1}).

Finally, as \citet{2006ApJ...638..759S} argue, supernova progenitors,
such as RSGs and Luminous Blue Variables, may produce
significant amount of crystalline silicates
\citep[e.g.~][]{MWT_99_afgl4106,VWM_99_exgalxsil}, and thus contribute
to the interstellar crystallinity. However, the subsequent supernova
shocks travelling through these earlier ejecta will destroy a large
fraction of the grains by sputtering
\citep{JTH_94_graindestruction,JTH_96_grainsize}. The crystalline
and amorphous silicate materials (assumed to be affected in equal
amounts) will be returned to the gas phase. {Thus,} the ISM will be
enriched by the atomic building blocks of silicates, but in
gaseous form. Moreover, the high supernova rate in starburst galaxies
will affect the dust destruction rates $k_3$ ($=k_4$), 
returning {even} more dust to the gas-phase.

\subsubsection{Dust formation in the ISM}

{The (re-)condensation of dust from the gas phase in the ISM will 
also affect the crystallinity}. In our model calculations, we
assumed that the dust formation in the ISM itself is negligible as a
source of dust, hence $F_\mathrm{A} = F_\mathrm{X} = 0$. However,
calculations by \citet{2008A&A...479..453Z} show that it can be the
most important process that contributes to the dust content of the
Milky Way. Since dust formation and dust growth at the low
temperatures prevalent in molecular clouds yields only amorphous
silicates, inclusion of this factor will further reduce the
crystallinity of interstellar silicates.

\subsection{Low crystallinities in typical starbursts}

Although our models are successful in explaining the crystallinity
observed in ULIRGs \citep{2006ApJ...638..759S}, we find that this is
only the case for an extreme set of input parameters. For more usual
values for the star formation rate and the dust formation efficiency
in supernovae, and when the effect of supernovae on grain destruction
and amorphisation is taken into account, we find that the observed
crystallinity of silicates cannot be explained by a starburst driven
model only.  {Thus, we conclude that the high level of
  crystallinity derived by \citet{2006ApJ...638..759S} is inconsistent
  with the formation of crystalline silicates in stellar ejecta only.
  However their observations are also consistent with crystallinities
  a factor of $\sim$ 3--4.5 lower than what is reported in their paper
  when different resonances are used, in which case our model
  calculations show that stellar sources could indeed explain the
  observed crystallinity, using reasonable parameters. }

\section{Conclusion}

{In order to explain the high crystalline fraction of 
silicates in ULIRGs, as reported by} \citet{2006ApJ...638..759S} we
{find} that an additional source
of crystalline silicates (or crystallisation of
amorphous silicates) must be present, related to the AGN itself,
rather than the starburst activity.  {In this case, the
  crystallisation will occur due to heating by (UV) photons from the
  AGN environment, rather than (massive) stars. A potential scenario
  may include the formation of (crystalline) silicates in quasar winds
  \citep{2002ApJ...567L.107E}.} {In addition, further observational
studies are useful to better establish the crystalline fraction
of silicates in ULIRGs, and validate the conclusions presented here.}

\section*{Acknowledgements}

We thank Sacha Hony and Svitlana Zhukovska for careful reading of the
manuscript. Their comments and suggestions have led to a considerable
improvement of this work. This research benefited from the award of 
a Leverhulme Research Fellowship to F.~Kemper. 



\end{document}